# Mid-infrared GeSn electro-absorption optical modulators on silicon

Jun-Han Lin,[1] Bo-Jung Huang,[1] H. H. Cheng,[2] and Guo-En Chang[1,*]

[1]Department of Mechanical Engineering, and Advanced Institute of Manufacturing with High-tech Innovations, National Chung Cheng University, Chiayi County 62102, Taiwan
[2]Center for Condensed Matter Sciences, and Graduate Institute of Electronics Engineering, National Taiwan University, Taipei 10617, Taiwan

## Abstract

Mid-infrared silicon photonics has recently emerged as a new technology for a wide range of applications such as optical communication, lidar, and bio-sensing. One key component enabling this technology is the mid-infrared optical modulator used for encoding optical signals. Here, we present a GeSn electro-absorption modulator that can operate in the mid-infrared range. Importantly, this device is monolithically integrated on a silicon substrate, which provides compatibility with standard complementary metal-oxide-semiconductor technology for scalable manufacturing. By alloying Ge with Sn to engineer the bandgap, we observed a clear Franz–Keldysh effect and achieved optimal modulation in the mid-infrared range of 2067–2208 nm with a maximum absorption ratio of 1.8. The results on the Si-based mid-infrared optical modulator open a new avenue for next-generation mid-infrared silicon photonics.

* correspondence and material requests should be addressed to G.E.C. (email:imegec@ccu.edu.tw)

**Introduction**

The operation wavelength of silicon photonics has been recently extended from the near-infrared (NIR) to the mid-infrared (MIR) spectral region (typically defined as a wavelength range of 1.8–5 μm [1]) for many emerging applications such as ultra high-speed optical communications, remote sensing, infrared version, and lidar [1–5]. With the unique advantages of complementary metal-oxide-semiconductor (CMOS) compatibility, various electronic and photonic devices can be monolithically integrated on the same chip to realize an on-chip MIR photonic system. However, on-chip MIR photonic systems rely on both active and passive fundamental devices that can operate in the MIR spectral region, and consequently, extensive efforts have been made to develop MIR photonic devices on Si. Recently, passive MIR devices such as low-loss waveguides, couplers, and splitters have been demonstrated [6]. In addition, active group-IV MIR devices including lasers [7,8] and photodetectors [9] have been demonstrated recently, bringing MIR silicon photonics closer to widespread applications.

An optical modulator is a crucial component used for encoding optical signals in any photonic system. However, efficient MIR optical modulators based on group-IV semiconductors pose fundamental and scientific challenges. First, the symmetry of crystalline Si and Ge leads to vanishing linear electro-optic (Pockels) effects for optical modulators. Free-carrier plasma dispersion effects have been employed to fabricate high-speed Si optical modulators [10–12]. However, relatively weak plasma dispersion effects require a long device length (typically several hundred micrometers or longer) to achieve efficient optical modulation, which pose challenges to high-density integration in Si photonics. Alternatively, introducing resonators with high quality factors can soften the device footprint requirements, but at the price of a significantly narrowed optical bandwidth of only a few nanometers. There is, fortunately, a third option that could potentially be developed into a competitive solution for efficient group-IV optical modulators based on electro-absorption (EA) effect in which the optical absorption coefficient in semiconductors can be modified via applications of electric fields. Although Ge is an indirect bandgap semiconductor, it surprisingly exhibits efficient electro-absorption effects associated with direct-gap transitions at strengths comparable to III-V direct bandgap semiconductors, including the Franz–Keldysh (FK) effect in bulk Ge and GeSi [13] and the quantum-confined Stark effect (QCSE) in Ge quantum wells [14]. With efficient, ultrafast EA effects, high-performance Ge-based electro-absorption modulators (EAMs) have been demonstrated [15–17]. Unfortunately, the relatively large direct bandgap in Ge dictates that Ge-based EAMs can only operate near 1500-nm wavelength range, not in the MIR range. As a result, efficient MIR optical modulators based on group-IV materials remain elusive for completing MIR photonic systems on Si.

Recently, GeSn alloys have emerged as a new class of group-IV semiconductors for MIR Si photonics because of the tunability of their band structure, which can be tuned from indirect to direct, their bandgap energy, which can be reduced from that of Ge, and their CMOS compatibility. These advantages have led to the development of GeSn-based active photonic devices including light emitters [7,8] and photodetectors [18-22] capable of operating in the MIR region. While there have been some reports studying the electro-absorption FK effects in GeSn [23,24], to date, GeSn-based optical modulators have not been experimentally realized. In this study, we demonstrate, to the best of our knowledge, the first GeSn EAM on silicon based on the FK effect. By introducing Sn into the active layer to engineer the direct bandgap energy, the absorption edge is extended to the MIR region. Upon the application of an electric field, we observe a clear FK effect and achieve optimal optical modulation in the MIR range, confirming the feasibility of GeSn alloys for efficient MIR optical modulators on Si.

# Results

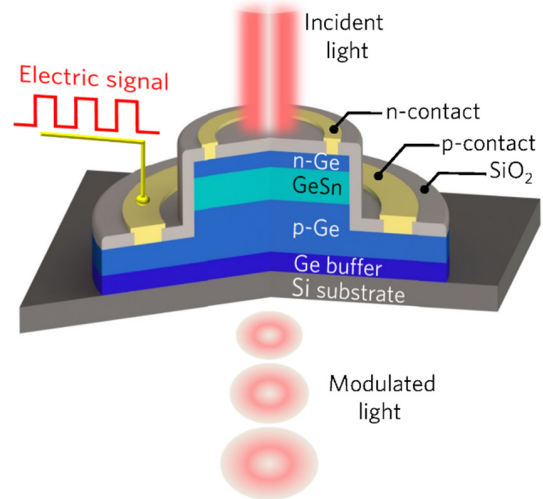

**Figure 1.** Schematic of our designed normal-incident GeSn optical modulator. The mid-infrared light beam is incident from the top and transmitted through the GeSn active layer. Bias voltage is applied to the *p-i-n* junction to modulate the MIR light based on the FK effect.

**Design of GeSn electro-absorption modulators.** Figure 1 shows a schematic diagram of the designed and fabricated normal-incident GeSn EAM grown using low-temperature molecular beam epitaxy (MBE). The device consists of a pseudomorphic Ge/GeSn/Ge *p-i-n* junction with a thickness of 140/360/390 nm on a double-side polished Si(001) substrate via a fully strain-relaxed Ge virtual substrate (thickness 120 nm). The GeSn active layer has a Sn composition of 5.2% and a compressive strain of 0.74%, which were confirmed by x-ray and calibrated secondary ion mass spectrometry measurements. The samples were fabricated into a normal incident EAM using CMOS-compatible processing technology [21] with a circular mesa (D) of 500 μm, a $SiO_2$ passivation layer with a thickness of 400 nm, and Cr/Au metal pads with a thickness of 20/200 nm. As light is normally incident on the surface of the GeSn EAM, the EA effect takes place when an electric field is applied to the GeSn active layer under the revised bias condition, thereby modifying the absorption coefficient near the direct bandgap. As a result, the intensity of the transmitted light can be modulated.

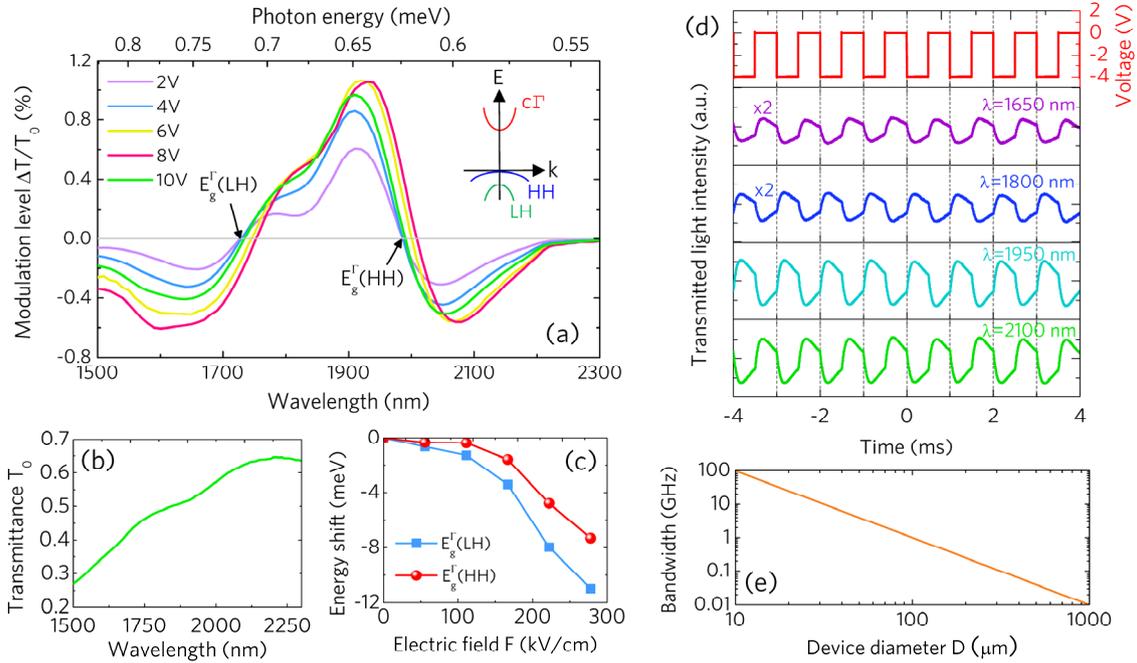

**Figure 2.** (a) Modulation level spectra of the GeSn EAM measured at different swing voltages. Clear variation of transmittance is observed, showing the FK effect. The inset shows the schematic band structure of the GeSn active layer. The direct-gap transition energies are indicated by arrows. (b) Transmittance spectrum of the device under zero-bias condition. (c) Shifts of the direct-gap transition energies as a function of the applied electric field. (d) AC optical modulation at selected wavelengths with a swing voltage of 0 to -4V at 1 kHz. (e) Calculated EAM bandwidth as a function of device diameter.

**Optical modulation characteristics.** Figure 2(a) shows the measured modulation level, defined as $\Delta T(V_s)/T_0 = [T(V_s) - T_0]/T_0$, where $T_0$ and $T(V_s)$ are the transmittance measured at zero bias (Fig. 2(b)) and at a swing voltage of $V_s$ starting from 0 V, respectively. The spectra clearly show significant oscillation characteristics in the range of 1500–2200 nm, and the variation of the transmittance increases with increasing swing voltage. In addition, the transmittance spectra exhibit redshift as the swing voltage increases because of the electric-field-induced tilting of the energy bands. These observations clearly demonstrate the FK effect in the GeSn active layer and electrically controlled optical modulation in the MIR region for the GeSn EAM. From the spectra, two main oscillation characteristics were observed around 1730 nm and 2000 nm, which are associated with direct optical transitions from the light-hole band to the conduction band ( $\text{LH} \rightarrow c\Gamma$ ) and from the heavy-hole band to the conduction band ( $\text{HH} \rightarrow c\Gamma$ ), respectively. The transition energies were determined to be $E_g^\Gamma(\text{LH}) = 0.715\,\text{eV}$ and $E_g^\Gamma(\text{HH}) = 0.624\,\text{eV}$ from $\Delta T(V_s)/T_0 = 0$. The splitting of the HH and LH bands is caused by the compressive strain in the GeSn active layer (discussed later).

Figure 2(b) shows the transmittance spectrum of the GeSn EAM under the zero-bias condition. Figure 2(c) reveals the direct transition energy shift as a function of the applied electric field extracted from Fig. 2(a). The result indicates that the transition energy shift for both the $\text{LH} \rightarrow c\Gamma$ and $\text{HH} \rightarrow c\Gamma$ transitions increases with the applied electric field. The FK shift for the $\text{LH} \rightarrow c\Gamma$ transition is greater than that of the $\text{HH} \rightarrow c\Gamma$ transition, which can be primarily attributed to the smaller effective mass of LH than that of HH. From the modulation level spectra in Fig 2(a), high modulation levels of 1.05% at 1918 nm and 0.57% at 2066 nm were obtained, corresponding to extinction ratios of $0.127\,\text{dB}/\mu\text{m}$ and $0.688\,\text{dB}/\mu\text{m}$, respectively.

To demonstrate the modulation capacity in the MIR region, the AC modulation signal was obtained from the GeSn EAM at 1 kHz acquired by an oscillator with $V_s = 4\,V$, as shown in Fig. 2(d), where clear dynamical modulation is observed. For EAMs, the modulation speed is limited by the RC time delay and can be improved by reducing the device footprint [22], thus achieving high-speed MIR optical modulation. We then estimated the bandwidth of the GeSn EAM as a function of the device diameter using

$$f_{RC} = \frac{1}{2\pi RC} = \frac{1}{2\pi \varepsilon R} \frac{t}{A} \tag{1}$$

where $C$ is the intrinsic depletion capacitance, $\varepsilon$ is the permittivity, $R$ is the resistance, and $A$ and $t$ are the area and the thickness of the GeSn active layer, respectively. By choosing a standard load resistance of $R$=50 $\Omega$, which is widely used in microwave measured systems, the bandwidth of the GeSn EAM as a function of the device diameter was calculated, and the results are depicted in Fig. 2(e). The results show that the modulation speed can be significantly enhanced by decreasing the device diameter, and a high modulation speed of $f_{RC}$>10 GHz is achievable for D<30 μm.

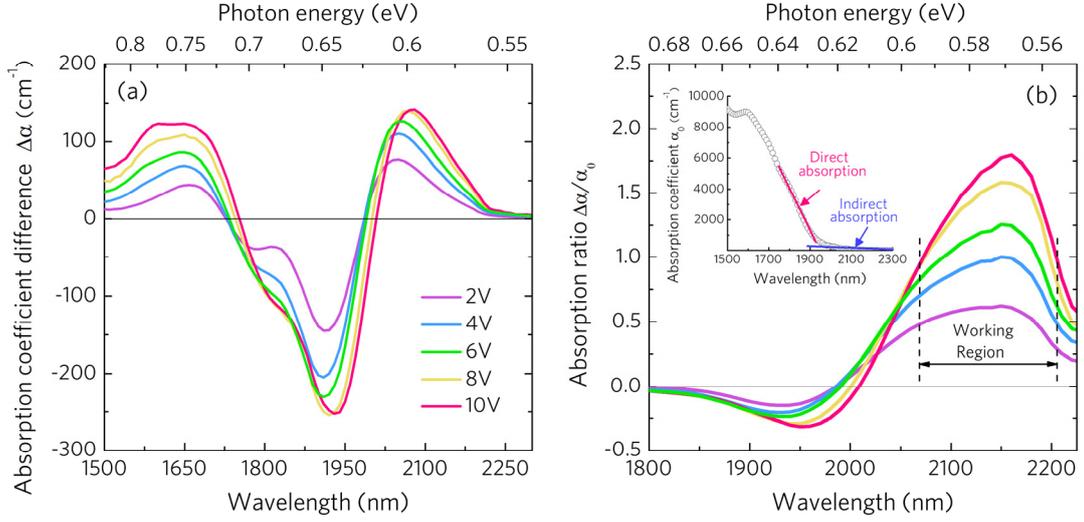

**Figure 3.** (a) Change in the absorption coefficient of the GeSn EAM at different swing voltages. (b) Absorption ratio of the GeSn active layer at different swing voltages. The inset shows the absorption coefficient spectrum under the zero-bias condition.

**Franz-Keldysh effect in GeSn.** To analyze the FK effect, the absorption coefficient change ($\Delta\alpha$) with different swing voltages for the GeSn active layer was extracted from the modulation level spectra using [25]

$$\Delta\alpha = -\frac{1}{t}\ln\left[1+\frac{\Delta T(V_s)}{T_0}\right] \qquad (2)$$

where *t* is the thickness of the GeSn active layer, and the result is depicted in Fig. 3(a). The $\Delta\alpha$ spectra clearly show that the magnitude increases with the swing voltage, but the magnitude tends to saturate for $V_s > 6\ V$ with significant changes in the absorption coefficient occurring at $V_s = 10\text{V}$, e.g., 254 cm$^{-1}$ at 1923 nm and 140.61 cm$^{-1}$ at 2068 nm. These values are smaller than those obtained from Ge [13], mainly because of the strain-induced HH-LH splitting that weakens the oscillator strength in the FK effect near the absorption edge.

While a high extinction ratio is favorable for all EAMs, a low insertion loss is equally essential. Thus, the absorption ratio, $\Delta\alpha/\alpha_0$, with $\alpha_0$ being the absorption coefficient at zero bias, is usually used as the figure-of-merit for characterizing EAMs and for determining the optimal working region. Figure 3(b) depicts the absorption ratio spectra with different swing voltages for the GeSn active layer, where the inset shows the $\alpha_0$ spectrum experimentally obtained from the responsivity measurement (Supporting Information). The result shows that $\Delta\alpha/\alpha_0$ increases with higher swing voltages. The maximum absorption ratio of 1.8 was obtained at 2160 nm with $V_s = 10\ V$, which is equivalent to a scaling performance of a 3.01-dB extinction ratio with a 1.67-dB insertion loss for a 100-μm-long device. Despite the smaller $\Delta\alpha$, the obtained $\Delta\alpha/\alpha_0$ value is comparable with, or even better than, Ge-based EAMs [17, 26]. This behavior can be explained by the contribution of the indirect transition to the absorption coefficient. In pure Ge, the L-conduction band lies below the $\Gamma$-conduction band by a significant energy separation of $E_{\Gamma L} = 136\ \text{meV}$. As a result, there is usually a long indirect absorption tail extending below the direct transition band edge, which contributes to a permanent high absorption coefficient ($\alpha_L \propto (\hbar\omega - E_g^L)^2$, where $E_g^L$ is the indirect bandgap [27]) near the direct-gap absorption edge. Consequently, Ge EAMs usually suffer from relatively high insertion loss and a limited absorption ratio. In our GeSn EAM, however, the energy difference $E_{\Gamma L}$ between the two valleys is considerably reduced to 85 meV, leading to a significantly reduced $\alpha_L$ near the direct-gap absorption edge. As a result, the insertion loss is reduced, and the absorption ratio is increased. Across the wavelength range of 2067–2208 nm in Fig. 3(b), the absorption ratio is always greater than 1 for $V_s = 10\ V$. This result implies that the extinction ratio can indeed be greater than the insertion loss, which is beneficial for achieving high-performance optical modulation. We thus define this 141-nm-wide range as the operating regime of this device. Moreover, the optimal working regime of this GeSn EAM perfectly matches the emerging 2-μm MIR optical communication band [1, 28] and 2-μm lidar systems [29], making it ideally suitable for MIR optical communication applications.

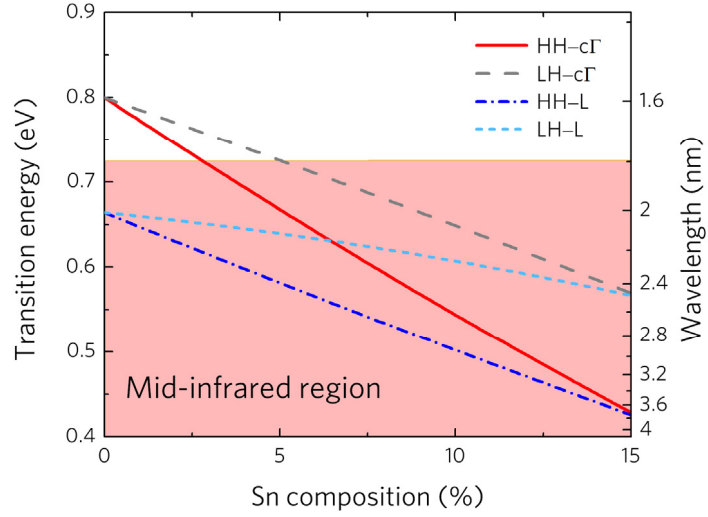

**Figure 4.** Calculated direct and indirect interband transition energies for pseudomorphic $Ge_{1-x}Sn_x$ on Ge.

**Theoretical modeling for modulation range.** The FK effect observed in GeSn in this study enables further development of MIR EAMs on Si. First, because the direct bandgap of $Ge_{1-x}Sn_x$ can be controlled by adjusting the Sn concentration, the working region of GeSn-based EAMs is expected to be "tunable" in the MIR region for a wide range of applications. To investigate the effects of Sn alloying, we calculated the direct bandgap energy of pseudomorphic $Ge_{1-x}Sn_x$ on Ge considering the strain effects using the model solid theory and deformation potential theory [25, 30] (Supporting Information), and the results are depicted in Fig. 4. Despite the compressive strain ($\varepsilon = -0.143x$), the direct bandgap energy can be considerably reduced with increasing Sn content. In addition, the compressive strain splits the valence band, pushing the HH band above the LH band. As a result, the HH→c$\Gamma$ transition is the lowest transition that defines the direct-gap absorption edge of the GeSn active layer, which is the working region of the GeSn EAM. The calculation indicates that an Sn content of 2.8% is required to redshift the direct absorption edge to the MIR range for optical modulation. Up to now, GeSn alloys with Sn compositions >15% have been experimentally demonstrated [8]. Therefore, with a Sn composition range of 2.8–15%, optical modulation in the broad spectral range of 1.8–4.26 μm can be achieved for a wide range of applications.

**Discussion.** In summary, we have experimentally demonstrated the first GeSn MIR optical modulator on Si. The Franz–Keldysh effect in MIR region was clearly observed. The spectral response can be tuned through bandgap engineering the GeSn alloy by incorporating various amounts of Sn into the active layer. At an Sn composition of 5.2%, optimal optical modulation in the MIR region of 2067–2208 nm was achieved with an absorption ratio up to 1.8. In addition, the ability to turn the modulation range by adjusting the Sn composition should enable efficient GeSn electro-absorption modulators operating at different mid-infrared spectral ranges for a wide range of applications. We also anticipate that the GeSn/Ge heterostructures can be used to develop waveguide optical modulators for MIR silicon photonic integrated circuits. Our demonstration of GeSn mid-infrared electro-absorption optical modulators represent a remarkable step in the development of efficient Si-based optical modulators, opening up new avenues for large-scale MIR Si photonics based on CMOS compatible technology.

**Methods**

**Sample growth.** The samples used in this study were grown on p-type double-side polished Si(001) substrates using molecular beam epitaxy at a base pressure of less than $2\times10^{-10}$ torr. The first step of the epitaxy process was the growth of a fully strain-relaxed Ge virtual substrate (VS) using a two-step growth technique, including a 100-nm-thick Si layer grown at 650°C, a 100-nm-thick Si layer grown at 350°C, a 60-nm-thick seed Ge layer grown at 350°C, followed by in-situ annealing at 800°C for 5 min, and a 60-nm-thick Ge buffer layer grown at 550°C. Then, a 390-nm-thick B-doped p-type Ge layer was grown at 550°C. The growth temperature was subsequently decreased to 150°C for the growth of the 360-nm-thick GeSn layer, followed by the growth of a 190-nm-thick Sb-doped n-type Ge layer. The epitaxy growth was completed with the growth of a 3-nm-thick Si cap layer.

**Transmission experiments.** The transmission experiments were carried out at room temperature using a broad-band quartz-tungsten-halogen (QTH) lamp as the light source, which was filtered using a 1200-nm high-pass filter, and dispersed using a monochromator equipped with a 600-line/mm grating blazed at 1600 nm. The dispersed light was then reshaped using an adjustable aperture and focused onto the top surface of the device through a 20X objective; this process ensured that the beam size was smaller than the window of the device. A square-wave AC bias with a frequency of 1 KHz was applied on the device using a wavefunction generator. The light transmitted through the device was then focused on a LN2-cooled InSb photodetector (1–5 µm detection range) and converted to an electrical signal, which was read out using a lock-in amplifier to determine the transmittance.

**Data availability.** The data that support the finding of this work are available from the corresponding author upon request.

**REFERENCES**

1. Soref, R. Mid-infrared photonics. In *Optical Fiber Communication Conference*, OSA Technical Digest (online); Optical Society of America, 2015, paper W4A.4.
2. Liu, J.; Cannon, D. D.; Wada, K.; Ishikawa, Y.; Jongthammanurak, S.; Danielson, D. T.; Michel, J.; Kimerling, L. C. Mid-infrared photonics in silicon and germanium. *Nat. Photon.* **2010**, *4*, 495-497.
3. Hu, T.; Dong, B.; Luo, X.; Liow, T.-Y.; Song, J.; Lee, C.; Lo, G.-Q. Silicon photonic platforms for mid-infrared applications. *Photon. Res.* **2017**, *5*, 417-430.
4. Refaat, T. F.; Ismail, S.; Koch, G. J.; Rubio, M.; Mack, T. L.; Notari, A.; Collins, J. E.; Lewis, J.; Young, R. D.; Choi, Y.; Abedin, M. N.; Singh, U. N. Backscatter 2-µm Lidar validation for atmospheric $CO_2$ differential absorption Lidar applications. *IEEE Trans. Geosci. Remote Sensing* **2011**, *49*, 572-580.
5. Chang, C.; Li, H.; Ku, C.-T.; Yang, S.-G.; Cheng, H. H.; Hendrickson, J.; Soref, R. A.; Sun, G. 1.6–1.9 µm infrared vision. Appl. Opt. **55**, 10170 (2016).
6. Nedeljkovic, M.; Khokhar, A. Z.; Hu, Y.; Chen, X.; Penades, J. S.; Stankovic, S.; Chong, H. M. H.; Thomson, D. J.; Gardes, F. Y.; Reed, G. T.; Mashanovich, G. Z. Silicon photonic devices and platforms for the mid-infrared. *Opt. Mater. Express* **2013**, *3*, 1205-1214.
7. Wirths, S.; Geiger, R.; von den Driesch, N.; Mussler, G.; Stoica, T.; Mantl, S.; Ikonic, Z.;

Luysberg, M.; Chiussi, S.; Hartmann, J. M.; Sigg, H.; Faist, J.; Buca, D.; Grutzmacher, D. Lasing in direct-bandgap GeSn alloy. *Nat. Photon.* **2015**, *9*, 88-92.

8. Margetis, J.; Al-Kabi, S.; Du, W.; Dou, W.; Zhou, Y.; Pham, T.; Grant, P.; Ghetmiri, S.; Mosleh, A.; Li, B.; Liu, J.; Sun, G.; Soref, R.; Tolle, J.; Mortazavi, M.; Yu, S.-Q. Si-Based GeSn lasers with wavelength coverage of 2-3 μm and operating temperatures up to 180 K. *ACS Photonics* **2018**, *5*, 827-833.

9. Ackert, J. J.; Thomson, D. J.; Shen, L.; Peacock, A. C.; Jessop, P. E.; Reed, G. T.; Mashanovich, G. Z.; Knights, A. P. High-speed detection at two micrometres with monolithic silicon photodiodes. *Nat. Photon.* **2015**, *9*, 393-396.

10. Liu, A.; Jones, R.; Liao, L.; Samara-Rubio, D.; Rubin, D.; Cohen, O.; Nicolaescu, R.; Paniccia, M. A high-speed silicon optical modulator based on a metal–oxide–semiconductor capacitor. *Nature* **2004**, *427*, 615-618.

11. Reed, G. T.; Masganovish, G.; Gardes, F. Y.; Thomson, D. J. Silicon optical modulators. *Nat. Photon.* **2010,** *4*, 518-526.

12. Camp, M. A. V.; Assefa, S.; Gill, D. M.; Barwicz, T.; Shank, S. M.; Rice, P. M.; Topuria, T.; Green, W. M. J. Demonstration of electrooptic modulation at 2165nm using a silicon Mach-Zehnder interferometer. *Opt. Express* **2012**, *20*, 28009-28016.

13. Jongthammanurak, S.; Liu, J.; Wada, K.; Cannon, D. D.; Danielson, D. T.; Pan, D.; Kimerling, L. C.; Michel, J. Large electro-optic effect in tensile strained Ge-on-Si films. *Appl. Phys. Lett.* **2006**, 89, 161115.

14. Kuo, Y.-H.; Lee, Y. K.; Ge, Y.; Ren, S.; Roth, J. E.; Kamins, T. I.; Miller, D. A. B.; Harris, J. S. Strong quantum-confined Stark effect in germanium quantum-well structures on silicon. *Nature* **2005**, 437, 1334-1336.

15. Liu, J.; Beals, M.; Pomerene, A.; Bernardis, S.; Sun, R.; Cheng, J.; Kimerling, L. C.; Michel, J. Waveguide-integrated, ultralow-energy GeSi electro-absorption modulators. *Nat. Photon*. **2008**, 2, 433-437.

16. Chaisakul, P.; Marris-Morini, D.; Rouifed, M.-S.; Isella, G.; Chrastina, D.; Frigerio, J.; Roux, X. L.; Edmond, S.; Coudevylle, J.-R.; Vivien, L. 23 GHz Ge/SiGe multiple quantum well electro-absorption modulator. *Opt. Express* **2012**, *20*, 3219-3224.

17. Feng, D.; Liao, S.; Liang, H.; Fong, J.; Bijlani, B.; Shafiiha, R.; Luff, B. J.; Luo, Y.; Cunningham, J.; Krishnamoorthy, A. V.; Asghari, M. High speed GeSi electro-absorption modulator at 1550 nm wavelength on SOI waveguide. *Opt. Express* **2012**, *20*, 22224-22232.

18. Oehme, M.; Schmid, M.; Kaschel, M.; Gollhofer, M.; Widmann, D.; Kasper, E.; Schulze, J. GeSn p-i-n detectors integrated on Si with up to 4% Sn. *Appl. Phys. Lett.* **2012**, *101*, 141110.

19. Tseng, H. H.; Li, H.; Mashanov, V.; Yang, Y. J.; Cheng, H. H.; Chang, G. E.; Soref, R. A.; Sun, G. GeSn-based p-i-n photodiodes with strained active layer on a Si wafer. *Appl. Phys. Lett.* **2013**, *103*, 231907.

20. Pham, T.; Du, W.; Tran, H.; Margetis, J.; Tolle, J.; Sun, G.; Soref, R. A.; Naseem, H. A.; Li, B.; Yu, S.-Q. Systematic study of Si-based GeSn photodiodes with 2.6 μm detector cutoff for short-wave infrared detection. *Opt. Express* **2016**, *24*, 4519-4531.

21. Huang, B.-J.; Lin, J.-H.; Cheng, H. H.; Chang, G.-E. GeSn resonant-cavity-enhanced photodetectors on silicon-on-insulator platforms. *Opt. Lett.* **2018**, *43*, 1215-1218.

22. Oehme, M.; Kostecki, K.; Ye, K.; Bechler, S.; Ulbricht, K.; Schmid, M.; Kaschel, M.; Gollhofer, M.; Körner, R.; Zhang, W.; Kasper, E.; Schulze, J. GeSn-on-Si normal incidence photodetectors with bandwidths more than 40 GHz. *Opt. Express* **2014**, *22*, 839-846.

23. Soref, R. A.; Sun, G.; Cheng, H. H. Franz-Keldysh electro-absorption modulation in germanium-tin alloys. *J. Appl. Phys.* **2012**, *111*, 123113.

24. Oehme, M.; Kostecki, K.; Schmid, M.; Kaschel, M.; Gollhofer, M.; Ye, K.; Widmann, D.; Koerner, R.; Bechler, S.; Kasper, E.; Schulze, J. Franz-Keldysh effect in GeSn pin photodetectors. *Appl. Phys. Lett.* **2014**, *104*, 161115.

25. Chuang, S. L. *Physics of Photonic Devices*; Wiley: New York, 2009.

26. Mastronardi, L.; Banakar, M.; Khokhar, A.; Hattasan, N.; Rutirawut, T.; Bucio, T. D.; Grabska, K. M.; Littlejohns, C.; Bazin, A.; Mashanovich, G.; Gardes, F. High-speed Si/GeSi hetero-structure electro absorption modulator. *Opt. Express* **2018**, *26*, 6663-6673.

27. Pankove, J. I. *Optical Processes in Semiconductors*; Dover Publications: New York, 1971.

28. Chang, G. E.; Basu, R.; Mukhopadhyay, B.; Basu, P. K. Design and modeling of GeSn-based heterojunction phototransistors for communication applications. *IEEE J. Sel. Top. Quant. Electron.* **2016**, *22*, 425-433.

29. Refaat, T. F.; Ismail, S.; Koch, G. J.; Rubio, M.; Mack, T. L.; Notari, A.; Collins, J. E.; Lewis, J.; Young, R. D.; Choi, Y.; Abedin, M. N.; Singh, U. N. Backscatter 2-μm Lidar validation for atmospheric CO2 differential absorption Lidar applications. *IEEE Trans. Geosci. Remote Sens.* **2011**, *49*, 572-580.
30. Chang, G. E.; Chang, S. W.; Chuang, S. L. Strain-balanced $Ge_zSn_{1-z}$–$Si_xGe_ySn_{1-x-y}$ multiple-quantum-well lasers. *IEEE J. Quantum Electron.* **2010**, *46*, 1813–1820.

**ACKNOWLEDGMENTS**

This work at CCU was supported by the Ministry of Science and Technology, Taiwan (MOST) under project Nos. MOST 104-2923-E-194-003-MY3 and MOST 106-2628-E-194-003-MY2. The authors are grateful for Prof. Greg Sun at University of Massachusetts Boston for many insightful discussions. We also thank Dr. Hui Li at NTU and Yun-Da Hsieh at CCU for assistance in the experiments.

**Author contributions**

G.E.C. conceived the initial idea of this work and designed the devices. H.H.C designed and prepared the materials used in this study and carried out material characterizations. J.H.L. and B.J.H. fabricated the devices and performed electrical and optical measurements. G.E.C. performed the band structure calculations. J.H.L. and G.E.C. performed the data analysis. G.E.C. and H.H.C. wrote the manuscript. G.E.C. supervised the entire project.

**Additional Information**

**Supporting Information** accompanying this paper is available.
**Competing interests:** The authors declare no completing financial interests.

Supporting information for

# Mid-infrared GeSn Electro-Absorption Optical Modulators on Silicon

Jun-Han Lin,[1] Bo-Jun Huang,[1] H. H. Cheng,[2] and Guo-En Chang[*,1]

[1]Department of Mechanical Engineering, and Advanced Institute of Manufacturing with High-tech Innovations, National Chung Cheng University, Chiayi County 62102, Taiwan
[2]Center for Condensed Matter Sciences, and Graduate Institute of Electronics Engineering, National Taiwan University, Taipei 10617, Taiwan

**Responsivity experiments and absorption coefficient**

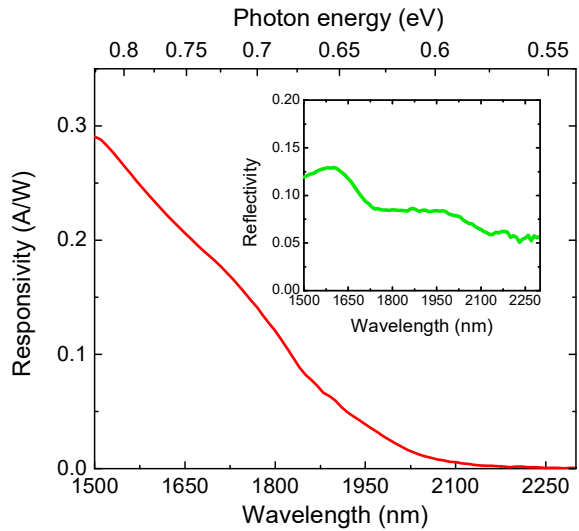

Figure S1. Room-temperature responsivity spectrum of the GeSn *p-i-n* diode measured at zero-bias conditions. The inset shows the measured reflectivity spectrum of the device.

Responsivity experiments were performed to obtain the zero-bias absorption coefficient ($a_0$) of the GeSn active layer. A QTH lamp was used as the light source, which was chopped at 200 Hz, filtered using a 1200-nm high-pass filter, dispersed using a monochromator equipped with a

600-line/mm grating blazed at 1600 nm, and then focused on the device in normally incident direction via a microscope. The generated photocurrent from the GeSn *p-i-n* diode under unbiased conditions was read out using a lock-in amplifier to enhance the signal-to-noise ratio for determining the responsivity ($R$). The reflected light from the device was collected using an extended-InGaAs photodetector for determining the reflectivity ($R_{refl}$). Figure S1 shows the measured zero-bias responsivity spectrum of the device, and the inset shows the measured reflectivity spectrum. From the responsivity and reflectivity spectra, the absorption coefficient of the GeSn active layer could be related to the responsivity by [1]

$$R = (1 - R_{refl})\frac{q\lambda}{hc}\exp(-\alpha_n t_n)[1 - \exp(-\alpha_i t_i)] \tag{S1}$$

where $q$ is the elementary charge, $\lambda$ is the free-space wavelength, $h$ is Planck's constant, $\alpha_n$ and $t_n$ are the absorption coefficient and thickness of the top *n*-type Ge layer, respectively, and $\alpha_i$ and $t_i$ are the absorption coefficient and thickness of the GeSn active layer, respectively. Taking the absorption coefficient of Ge from Ref. [2], the absorption coefficient of the GeSn layer was extracted using Eq. (S1).

**Band Structure Calculation**

The band structures were calculated using model-solid theory and deformation potential theory [3, 4]. The parameters for GeSn were obtained by linear interpolation between those of Ge and α-Sn, as listed in Table S1. The average valence band energy (in units of eV) for bulk $Ge_{1-x}Sn_x$ related to Ge can be written as

$$E_{vav}(Ge_{1-x}Sn_x) = 0.69x \tag{S2}$$

Then, the spin-orbital splitting energy can be expressed as

$$\Delta(Ge_{1-x}Sn_x) = 0.29 - 0.51x \tag{S3}$$

The heavy-hole (HH), light-hole (LH), and direct conduction band edges ($E_\Gamma$) under strain can then be determined as follows:

$$E_{HH} = \left(E_{vav} + \frac{\Delta}{3}\right) - P_\varepsilon - Q_\varepsilon \tag{S4}$$

$$E_{LH} = \left(E_{vav} + \frac{\Delta}{3}\right) - P_\varepsilon + \frac{1}{2}\left(Q_\varepsilon - \Delta + \sqrt{\Delta^2 + 2\Delta Q_\varepsilon + 9Q_\varepsilon^2}\right) \tag{S5}$$

$$E_\Gamma = \left(E_{vav} + \frac{\Delta}{3}\right) + E_g^\Gamma + P_{c,\varepsilon} \tag{S6}$$

where $P_\varepsilon$, $Q_\varepsilon$, and $P_{c,\varepsilon}$ are strain-induced energy shifts expressed as

$$P_\varepsilon = -a_v\left(\varepsilon_{xx} + \varepsilon_{yy} + \varepsilon_{zz}\right) \tag{S7}$$

$$P_{c,\varepsilon} = a_c\left(\varepsilon_{xx} + \varepsilon_{yy} + \varepsilon_{zz}\right) \tag{S8}$$

$$Q_\varepsilon = -\frac{b_v}{2}\left(\varepsilon_{xx} + \varepsilon_{yy} - 2\varepsilon_{zz}\right) \tag{S9}$$

In addition, $E_g^\Gamma$ is the direct bandgap energy for unstrained Ge$_{1-x}$Sn$_x$, and it can be expressed as

$$E_g^\Gamma(\text{Ge}_{1-x}\text{Sn}_x) = (1-x)E_g^\Gamma(\text{Ge}) + xE_g^\Gamma(\text{Sn}) - b^\Gamma x(1-x) \tag{S10}$$

where $E_g^\Gamma(\text{Ge})$ and $E_g^\Gamma(\text{Sn})$ are the direct bandgap energy of Ge and α-Sn, respectively, and $b^\Gamma = 2.42$ eV is the bowing energy [5]. In addition, $\varepsilon_{xx}$, $\varepsilon_{yy}$, and $\varepsilon_{zz}$ are the strains in the Ge$_{1-x}$Sn$_x$ layer. For pseudomorphic GeSn on Ge, the strain values can be calculated by

$$\varepsilon_{xx} = \varepsilon_{yy} = \frac{a(\text{Ge}) - a(\text{Ge}_{1-x}\text{Sn}_x)}{a(\text{Ge}_{1-x}\text{Sn}_x)} \tag{S11}$$

$$\varepsilon_{zz} = -2\frac{C_{12}}{C_{11}}\varepsilon_{xx} \tag{S12}$$

where $a(\text{Ge})$ and $a(\text{Ge}_{1-x}\text{Sn}_x)$ are the bulk lattice constant of Ge and Ge$_{1-x}$Sn$_x$, respectively. Here, $C_{11}$ and $C_{12}$ are the stiffness matrix elements of Ge$_{1-x}$Sn$_x$, and their ratio is [6]

$$\frac{C_{12}}{C_{11}} = 0.3738 + 0.1676x - 0.0296x^2 \tag{S13}$$

Table S1: Parameters of Ge and α-Sn [4]

| Parameters | Ge | α-Sn |
|---|---|---|
| Average valence band energy $E_{vav}$ (eV) | 0 | 0.69 |
| Spin-orbital splitting energy $\Delta$ (eV) | 0.29 | 0.8 |
| Direct bandgap energy $E_g^\Gamma$ (eV) | 0.8 | -0.413 |
| Deformation potential energy $a_c$ (eV) | -8.24 | -6.00 |
| Deformation potential energy $a_v$ (eV) | 1.24 | 1.58 |
| Deformation potential energy $b_v$ (eV) | -2.9 | -2.7 |
| Lattice constant $a$ (Å) | 5.6573 | 6.4892 |

**References:**

1. Oehme, M.; Widmann, D.; Kostecki, K.; Zaumseil, P.; Schwartz, B.; Gollhofer, M.; Koerner, R.; Bechler, S.; Kittler, M.; Kasper, E.; Schulze, J. GeSn/Ge multiquantum well photodetectors on Si substrates. *Opt. Lett.* **2014**, *39*, 4711-4714.
2. Palik, E. D. *Handbook of Optical Constants of Solids*; Academic: London, 1985.
3. Chuang, S. L. *Physics of Photonic Devices*; Wiley: New York, 2009.
4. Chang, G. E.; Chang, S. W.; Chuang, S. L. Strain-balanced $Ge_zSn_{1-z}$–$Si_xGe_ySn_{1-x-y}$ multiple-quantum-well lasers. *IEEE J. Quantum Electron.* **2010**, *46*, 1813-1820.
5. Lin, H.; Chen, R.; Lu, W; Huo, Y.; Kamins, T. I.; Harris, J. S. Investigation of the direct band gaps in $Ge_{1-x}Sn_x$ alloys with strain control by photoreflectance spectroscopy. *Appl. Phys. Lett.* **2012**, *100*, 102109.
6. Beeler, R; Roucka, R.; Chizmeshya, A. V. G.; Kouvetakis, J.; Menendez, J. Nonlinear structure-composition relationships in the $Ge_{1-y}Sn_y$ /Si(100) (y < 0.15) system. *Phys. Rev. B* **2011**, *84*, 035204.